# Quantum Discrete Adiabatic Linear Solver based on Block Encoding and Eigenvalue Separator

Guojian Wu[a], Fang Gao[a*], Qing Gao[b], Yu Pan[c]

[a] *School of Electrical Engineering, Guangxi University, Nanning, 530004, China*
[b] *Hangzhou Innovation Institute, Beihang University, Hangzhou 310051, China*
[c] *College of Control Science and Engineering, Zhejiang University, Hangzhou 310027, Zhejiang Province, China*

**Abstract**

Linear system solvers are widely used in scientific computing, with the primary goal of solving linear system problems. Classical iterative algorithms typically rely on the conjugate gradient method. The rise of quantum computing has spurred interest in quantum linear system problems (QLSP), particularly following the introduction of the HHL algorithm by Harrow et al. in 2009, which demonstrated the potential for exponential speedup compared to classical algorithms. However, the performance of the HHL algorithm is constrained by its dependence on the square of the condition number. To address this limitation, alternative approaches based on adiabatic quantum computing (AQC) have been proposed, which exhibit complexity scaling linearly with the condition number. AQC solves QLSP by smoothly varying the parameters of the Hamiltonian. However, this method suffers from high Hamiltonian simulation complexity. In response, this work designs new Hamiltonians and proposes a quantum discrete adiabatic linear solver based on block encoding and eigenvalue separation techniques (BEES-QDALS). This approach bypasses Hamiltonian simulation through a first-order approximation and leverages block encoding to achieve equivalent non-unitary operations on qubits. By comparing the fidelity of the original algorithm and BEES-QDALS when solving QLSP with a fixed number of steps, and the number of steps required to reach a target fidelity, it is found that BEES-QDALS significantly outperforms the original algorithm. Specifically, BEES-QDALS achieves higher fidelity with the same number of steps and requires fewer steps to reach the same target fidelity.

∗ Corresponding author
*E-mail address:* gjwu@st.gxu.edu.cn (G. Wu), fgao@gxu.edu.cn (F. Gao), gaoqing@buaa.edu.cn, ypan@zju.edu.cn

## 1. Introduction

Linear system solvers are extensively used in scientific computing [1], with the goal of solving linear system problems (LSP). Specifically, the objective is to solve the linear equation $Ax = b$, where $A$ is an $N$-dimensional complex matrix and $b$ is an $N$-dimensional complex vector. The most efficient classical iterative algorithm to date is based on the conjugate gradient method, with a time complexity of $O(N\sqrt{\kappa}\log(1/\epsilon))$, where $\kappa = \|A\|_2 \times \|A^{-1}\|_2$ is the condition number of the coefficient matrix $A$, $\|\cdot\|_2$ denotes matrix 2-norm, and $\epsilon$ is the target accuracy. With the rapid advancement of quantum computing, there has been growing interest in the quantum linear system problem (QLSP) [2-5], which seeks to efficiently compute $|x\rangle = A^{-1}|b\rangle / \|A^{-1}|b\rangle\|_2$ on a quantum computer, where $A$ is Hermitian, $|b\rangle$ is the normalized $b$ and $|x\rangle$ is the normalized $x$. One of the most notable quantum algorithms for solving QLSP is the HHL algorithm [6], proposed by Harrow et al. in 2009, which achieves a solution $|x\rangle$ with a complexity of $O(\log(N)\kappa^2/\varepsilon)$, offering potential exponential speedup over classical algorithms. Quantum algorithms for QLSP have a variety of applications, including electromagnetic scattering computation [7], solving differential equations [8,9], data fitting [10], machine learning [11,12], and more generally, solving partial differential equations [13].

It is worth noting that in the original HHL algorithm, the complexity scales quadratically with the condition number $\kappa$. To address this issue, recent work has proposed alternative methods based on adiabatic quantum computing (AQC) [14-17], where the complexity scales nearly linearly with $\kappa$ [18]. AQC is a general model of quantum computation that has been proven to be polynomially equivalent to the standard gate-based model in terms of time complexity. An AQC-based linear solver solves the QLSP by smoothly varying the parameters of the quantum system's Hamiltonian. According to the adiabatic theorem, if the relevant eigenstates remain non-degenerate throughout the process and the Hamiltonian changes sufficiently slowly, the evolving state will remain close to the eigenstate of the final Hamiltonian [19]. This eigenstate encodes the information needed to solve the problem.

Furthermore, Subaşı et al. explored the quantum circuit implementation of AQC-based linear solvers [20], resulting in a linear solver designed through quantum discrete adiabatic evolution [21], known as QDALS. It is important to note that Hamiltonian simulation is a key component of this algorithm. However, gate-based Hamiltonian simulation is associated with high computational complexity [22-24]. Therefore, exploring approaches that can reduce the complexity of Hamiltonian simulation in QDALS is a meaningful and valuable direction for further research.

To bypass the complex Hamiltonian simulation, in this work, we propose a quantum discrete adiabatic linear solver based on a block encoding technique (BE-based QDALS). We integrate an improved block encoding technique into QDALS, where the Hamiltonian simulation operator in the original QDALS is replaced with a first-order approximation. The block encoding technique allows us to implement equivalent non-unitary operations on qubits via block encoding operators, thus achieving discrete adiabatic evolution based on the first-order approximation of the Hamiltonian simulation operator to solve the QLSP. While the use of first-order approximations and the block encoding technique helps avoid the complexity of Hamiltonian simulation, the errors introduced by the first-order approximation prevent the use of BE-based QDALS from outputting correct solutions. To address this, we design an improved Hamiltonian, which ensures the feasibility of BE-based QDALS. However, compared to the original QDALS, the use of the improved Hamiltonian in BE-based QDALS increases the number of steps required for discrete adiabatic evolution and makes it challenging to achieve high-fidelity solutions. To overcome these challenges, we develop an Eigenvalue Separator that enlarges the gap between the eigenvalues of the Hamiltonian, enabling QLSP to be solved with fewer adiabatic evolution steps and higher fidelity. The effectiveness of the improved Hamiltonian and the Eigenvalue Separator is validated both theoretically and experimentally. We refer to this enhanced version of QDALS, incorporating the improved block encoding, improved Hamiltonian and Eigenvalue Separator, as BEES-QDALS.

The remainder of this paper is structured as follows: Sec. 2 provides a detailed introduction to BEES-QDALS, Sec. 3 presents a series of comprehensive case studies, and finally Sec. 4 concludes and suggests potential directions for future research.

## 2. BEES-QDALS

In the following, Sec. 2.1 provides a brief introduction to the original QDALS, Sec. 2.2 gives an overview of the overall framework of BEES-QDALS, Sec. 2.3 explains the improved Hamiltonian in detail, Sec. 2.4 covers the Eigenvalue Separator in depth, and Sec. 2.5 elaborates on the improved block encoding operator.

### *2.1. Original QDALS*

This subsection introduces the original QDALS [20]. For simplicity, we assume that the coefficient matrix $A$ in the QLSP is an $N$-dimensional invertible positive definite Hermitian matrix, with $n = \log_2 N$ being an integer. The adiabatic quantum computing-based algorithm for solving the QLSP centers on converting the QLSP into an equivalent eigenvalue problem, and then, within the AQC framework, preparing the quantum state $|x\rangle = A^{-1}|b\rangle / \left\| A^{-1}|b\rangle \right\|_2$.

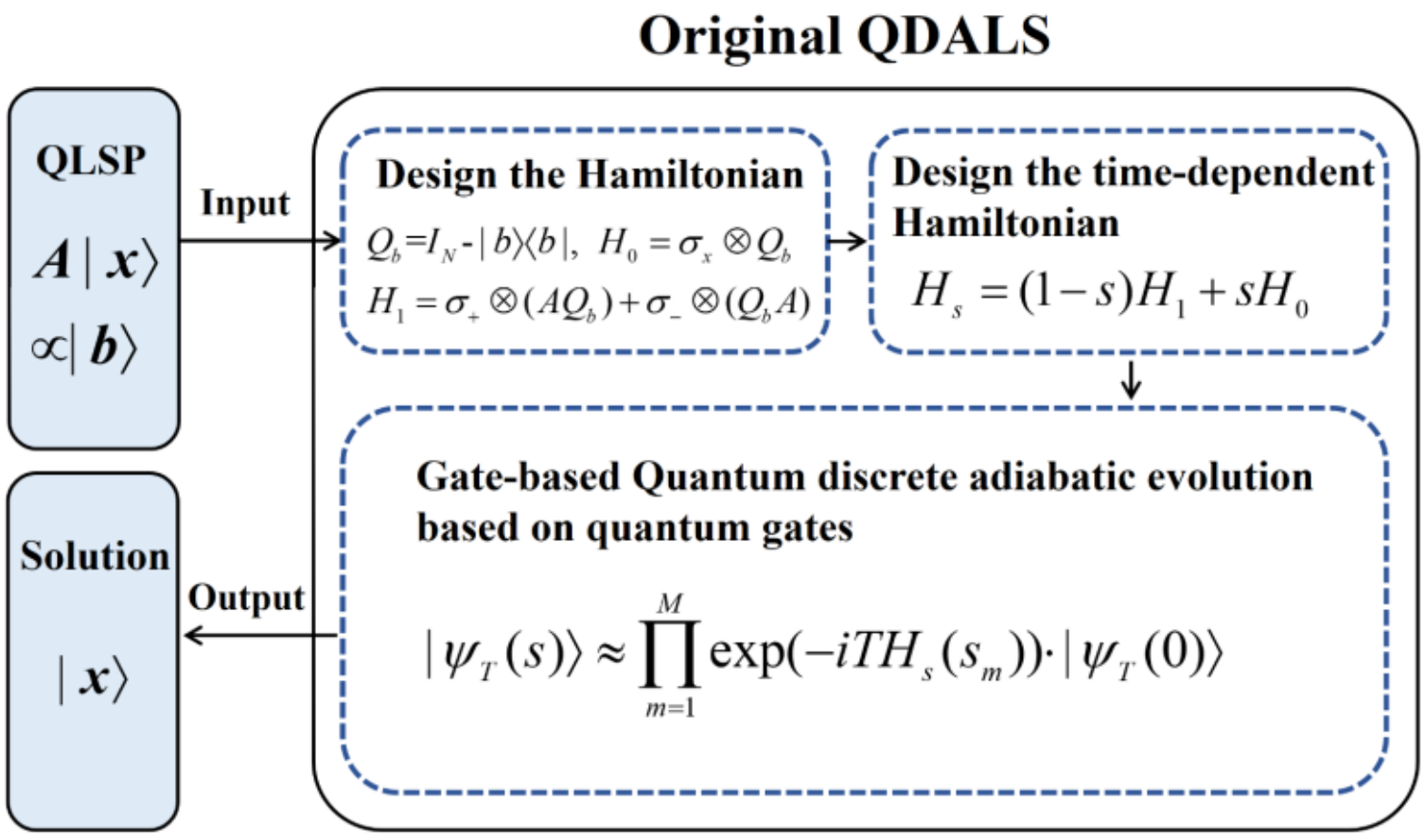

**Fig. 1.** Workflow of original QDALS

The workflow of the original QDALS is illustrated in **Fig. 1**, with the specific steps outlined as follows:

***Step 1*. Designing the Hamiltonian based on the known data from the QLSP problem.** With the known column vector $|b\rangle$, the projection operator is designed as

$$Q_b = I_N - |b\rangle\langle b|, \tag{1}$$

where $I_N$ represents the $N$-dimensional unit matrix. Then the initial Hamiltonian $H_0$ is designed as

$$H_0 = \sigma_x \otimes Q_b = \begin{bmatrix} 0 & 1 \\ 1 & 0 \end{bmatrix} \otimes Q_b = \begin{bmatrix} \boldsymbol{0}_{N\times N} & Q_b \\ Q_b & \boldsymbol{0}_{N\times N} \end{bmatrix}, \tag{2}$$

where $\boldsymbol{0}_{N\times N}$ represents the $N$-dimensional zero matrix. For the sake of convenience, two quantum states are defined as

$$|\boldsymbol{0}_N, b\rangle = \begin{bmatrix} \left.\begin{matrix} 0 \\ \vdots \\ 0 \end{matrix}\right\} N \\ b \end{bmatrix} = \begin{bmatrix} \boldsymbol{0}_N \\ b \end{bmatrix} \tag{3}$$

$$|b, \boldsymbol{0}_N\rangle = \begin{bmatrix} b \\ \left.\begin{matrix} 0 \\ \vdots \\ 0 \end{matrix}\right\} N \end{bmatrix} = \begin{bmatrix} b \\ \boldsymbol{0}_N \end{bmatrix}, \tag{4}$$

where $\boldsymbol{0}_N$ represents the $N$-dimensional zero column vector. Since

$$H_0 |\boldsymbol{0}_N, b\rangle = \begin{bmatrix} \boldsymbol{0}_{N\times N} & Q_b \\ Q_b & \boldsymbol{0}_{N\times N} \end{bmatrix} \begin{bmatrix} \boldsymbol{0}_N \\ b \end{bmatrix} = \boldsymbol{0}_{2N}, \tag{5}$$

$$H_0 |b, \boldsymbol{0}_N\rangle = \begin{bmatrix} \boldsymbol{0}_{N\times N} & Q_b \\ Q_b & \boldsymbol{0}_{N\times N} \end{bmatrix} \begin{bmatrix} b \\ \boldsymbol{0}_N \end{bmatrix} = \boldsymbol{0}_{2N}, \tag{6}$$

the null space of $H_0$ is spanned by $|\boldsymbol{0}_N, b\rangle$ and $|b, \boldsymbol{0}_N\rangle$. Then the final Hamiltonian $H_1$ is designed as

$$H_1 = \frac{1}{2}(\sigma_x + i\sigma_y) \otimes (AQ_b) + \frac{1}{2}(\sigma_x - i\sigma_y) \otimes (Q_b A) = \begin{bmatrix} \boldsymbol{0}_{N\times N} & AQ_b \\ Q_b A & \boldsymbol{0}_{N\times N} \end{bmatrix}. \tag{7}$$

When $A|x\rangle \propto |b\rangle$, there is

$$Q_b A |x\rangle = Q_b |b\rangle = \boldsymbol{0}_N. \tag{8}$$

Meanwhile, since

$$H_1 |x, \boldsymbol{0}_N\rangle = \begin{bmatrix} \boldsymbol{0}_{N\times N} & AQ_b \\ Q_b A & \boldsymbol{0}_{N\times N} \end{bmatrix} \begin{bmatrix} x \\ \boldsymbol{0}_N \end{bmatrix} = \boldsymbol{0}_{2N} \tag{9}$$

$$H_1 |\boldsymbol{0}_N, b\rangle = \begin{bmatrix} \boldsymbol{0}_{N\times N} & AQ_b \\ Q_b A & \boldsymbol{0}_{N\times N} \end{bmatrix} \begin{bmatrix} \boldsymbol{0}_N \\ b \end{bmatrix} = \boldsymbol{0}_{2N}, \tag{10}$$

the null space of $H_1$ is spanned by $|x, \boldsymbol{0}_N\rangle$ and $|\boldsymbol{0}_N, b\rangle$ 。

**Step 2. Designing the time-dependent Hamiltonian based on scheduling function.** For the scheduling function $f:[0,1] \to [0,1]$, $f(0)=0$, $f(1)=1$, we have

$$H_s(f(s)) = (1-f(s))H_0 + f(s)H_1, 0 \le s \le 1. \tag{11}$$

It's worth noting that since

$$H_s(f(s))|\boldsymbol{0}_N, b\rangle = (1-f(s))H_0|\boldsymbol{0}_N, b\rangle + f(s)H_1|\boldsymbol{0}_N, b\rangle = \boldsymbol{0}_{2N} \tag{12}$$

holds for any $s$, $|\boldsymbol{0}_N, b\rangle$ is always in the null space of $H_s(f(s))$. At the same time, there is a state $|\psi_T(f(s))\rangle$

that satisfies

$$H_s(f(s))\,|\,\psi_T(f(s)),\boldsymbol{0}_N\rangle = \boldsymbol{0}_{2N}\,, \tag{13}$$

which means the null space of $H_s(f(s))$ is spanned by $|\,\psi_T(f(s)),\boldsymbol{0}_N\rangle$ and $|\boldsymbol{0}_N, b\rangle$. Furthermore, since $|\,\psi_T(0),\boldsymbol{0}_N\rangle = |\,b,\boldsymbol{0}_N\rangle$ and $|\,\psi_T(1),\boldsymbol{0}_N\rangle = |\,x,\boldsymbol{0}_N\rangle$, $|\,\psi_T(f(s)),\boldsymbol{0}_N\rangle$ is the desired adiabatic path to solve QLSP [25], and there is

$$|\,\psi_T(f(s)),\boldsymbol{0}_N\rangle = \exp(-iT\int_0^s H_s(f(s'))ds')\,|\,\psi_T(0),\boldsymbol{0}_N\rangle\,. \tag{14}$$

In this work, the simplest scheduling function $f(s)=s$ is chosen, leading to the time-dependent Hamiltonian

$$H_s(s) = (1-s)H_0 + sH_1, 0 \le s \le 1\,. \tag{15}$$

**Step 3. Quantum discrete adiabatic evolution based on quantum gates.** To implement an efficient time-dependent Hamiltonian simulation of $H_s(s)$, a straightforward approach is to use the Trotter decomposition method, with the lowest-order approximation given by:

$$\begin{aligned}\exp(-iT\int_0^s H_s(s')ds') &\approx \prod_{m=1}^{M}\exp(-iThH_s(s_m)) \\ &\approx \prod_{m=1}^{M}\exp(-iTh(1-s_m)H_0)\exp(-iThs_mH_1)\end{aligned}\,, \tag{16}$$

where $h = s/M$, $s_m = mh$, with $h$ and $M$ viewed as the step size and the number of steps, respectively. The accuracy of quantum discrete adiabatic evolution will be higher with larger $M$ and smaller $h$. Specifying $\tau$ as a constant coefficient, the unitary operations $\exp(i\tau H_0)$ and $\exp(i\tau H_1)$ in Eq.(16) are the time-independent Hamiltonian simulation of the Hamiltonians $H_0$ and $H_1$. Here, we refer to the QDALS designed based on Hamiltonian simulation as HS-based QDALS.

When $s=1$, there is

$$\begin{aligned}|\,\psi_T(1),\boldsymbol{0}_N\rangle &= \exp(-iT\int_0^1 H_s(s')ds')\,|\,b,\boldsymbol{0}_N\rangle \\ &\approx \prod_{m=1}^{1/h}\exp(-iThH_s(s_m))\,|\,b,\boldsymbol{0}_N\rangle \approx |\,x,\boldsymbol{0}_N\rangle\end{aligned}\,, \tag{17}$$

which means the solution of the QLSP is stored in a quantum register and can be obtained by measurement.

## *2.2. Overview of BEES-QDALS*

This subsection provides an overview of BEES-QDALS, with detailed explanations of its key components presented sequentially in the following subsections.

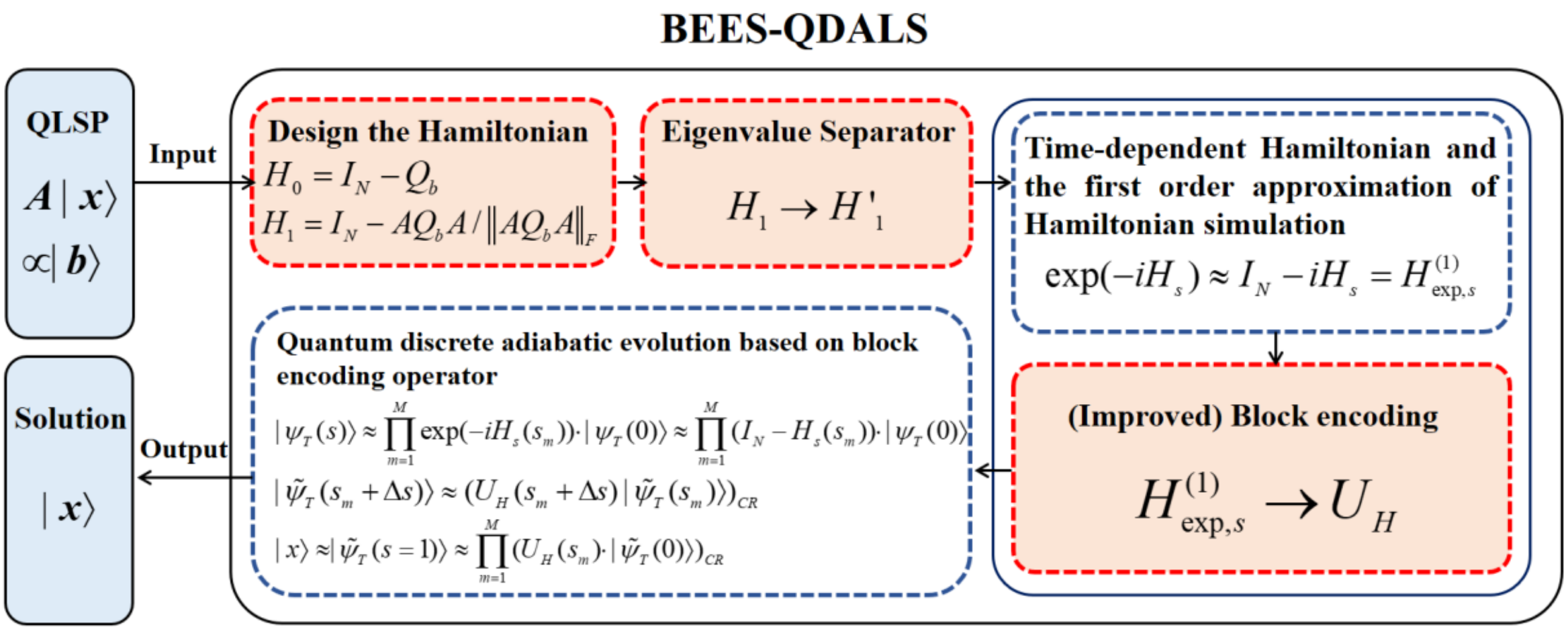

**Fig. 2.** Workflow of BEES-QDALS

The workflow of BEES-QDALS is shown in **Fig. 2**, the steps are as follows:

**Step 1. Designing the new initial Hamiltonian $H_0$ and final Hamiltonian $H_1$.** Based on the known data in the QLSP problem, $H_0$ and $H_1$ can be designed as:

$$H_0 = I_N - Q_b \, , \tag{18}$$

$$H_1 = I_N - \frac{AQ_bA}{\|AQ_bA\|_F} \, , \tag{19}$$

where $\|\cdot\|_F$ represents the Frobenius norm of the matrix, which can be obtained by the open-root calculation of the sum of the squares of the absolute values of the elements of the matrix. The specific properties of these two Hamiltonians will be given in Sec. 2.3.

**Step 2. Using Eigenvalue Separator to process $H_1$ to get $H'_1$.** The specific introduction of Eigenvalue Separator is given in Sec. 2.4.

**Step 3. Calculating the first-order approximation of Hamiltonian simulation.** We design the time-dependent Hamiltonian based on the initial Hamiltonian $H_0$ and the processed final Hamiltonian $H'_1$. By choosing the scheduling function as $f(s) = s$, we obtain $H_s(s) = (1-s)H_0 + sH'_1$. In this work, $H_s(s)$ is abbreviated as $H_s$. In order to avoid the high complexity of the Hamiltonian simulation operator $\exp(-iH_s)$, we take the first-order approximation of $\exp(-iH_s)$ as

$$\exp(-iH_s) \approx I_N - iH_s = H^{(1)}_{\exp,s} \, . \tag{20}$$

Here $H^{(1)}_{\exp,s}$ is usually non-unitary.

**Step 4. Block encoding.** Since $H^{(1)}_{\exp,s}$ is non-unitary, block encoding technology is used to achieve equivalent non-unitary operations on quantum states. With the block encoding technique, the quantum system requires the introduction of $n_A$ ancillary qubits. The entire quantum system consists of both the ancillary and main qubits. By using block encoding, we can construct the corresponding block-encoded quantum circuit, with the associated block encoding operator denoted as $U_H$ . A detailed explanation of this operator is provided in Sec. 2.5.

**Step 5. Quantum discrete adiabatic evolution based on block encoding operator.** For simplicity, the states of the entire quantum system and the qubits in the main quantum register at the discrete time point $s_m$ are denoted as $|\tilde{\psi}_T(s_m)\rangle$ and $|\psi_T(s_m)\rangle$ , respectively. The system state at the next discrete time point (i.e., $s_m+\Delta s$ ) is

$$|\tilde{\psi}_T(s_m+\Delta s)\rangle \approx (U_H(s_m+\Delta s)\,|\tilde{\psi}_T(s_m)\rangle)_{CR} = |\psi_T(s_m+\Delta s),\boldsymbol{0}_{2^{n_A}}\rangle\,, \tag{21}$$

where $(\;\cdot\;)_{CR}$ represents the erasure of redundant quantum states, that is, when the $n_A$ ancillary qubits collapse to $|0\rangle$ , the state of the main qubits is preserved. This can be achieved through a post-selection operation. Furthermore, by applying a series of block encoding operators and redundant quantum state erasure operations, we obtain:

$$|\tilde{x}\rangle \approx |\tilde{\psi}_T(s=1)\rangle \approx \prod_{m=1}^{M}(U_H(s_m)\cdot|\tilde{\psi}_T(0)\rangle)_{CR} = |x,\boldsymbol{0}_{2^{n_A}}\rangle\,, \tag{22}$$

where $\prod_{m=1}^{M}(\;\cdot\;)_{CR}$ represents the process of executing block encoding operators sequentially, where after each execution, redundant quantum states are erased. After a series of block encoding operators and redundant quantum state erasure operations, the state of the main qubits evolves to the solution of the QLSP.

### *2.3. Characteristics of the Hamiltonians in BEES-QDALS*

For the initial and final Hamiltonians in BEES-QDALS, we have the following lemmas:

***Lemma 1.*** For the initial Hamiltonian $H_0$ shown in Eq. (18), its largest eigenvalue is 1, with the corresponding eigenvector being $|b\rangle$ , while all other eigenvalues are 0.

***Proof.*** According to Eq. (1) and Eq. (18), we have

$$H_0 = |b\rangle\langle b|\,. \tag{23}$$

Obviously, $H_0$ is a space spanned by $|b\rangle$ , so we have

$$H_0\,|b\rangle = |b\rangle\langle b\,|\,b\rangle = |b\rangle\,. \tag{24}$$

The different eigenvalues of the Hermitian matrix $H_0$ correspond to orthogonal eigenvectors. We denote the eigenvector orthogonal to $|b\rangle$ as $|b^{\perp}\rangle$ . Since

$$H_0\,|b^{\perp}\rangle = |b\rangle\langle b\,|\,b^{\perp}\rangle = \boldsymbol{0}_N\,, \tag{25}$$

the other eigenvalues of the Hermitian matrix $H_0$ are 0.

In summary, the maximum eigenvalue of $H_0$ is 1, with the corresponding eigenvector being $|b\rangle$ , and the other eigenvalues are 0. The proof is complete.

***Lemma 2.*** For the final Hamiltonian $H_1$ in Eq. (19), its largest eigenvalue is 1, with the corresponding eigenvector being $|x\rangle$, while all the other eigenvalues belong to the interval [0, 1).

***Proof.*** The eigenvalues of matrix *A* are denoted as $\lambda_j^A$ for *j*=1,…, *N*, with corresponding eigenvectors $|u_j^A\rangle$. We define the projection coefficients of $|b\rangle$ onto various eigenvectors as

$$\beta_j = \langle u_j^A | b\rangle , \tag{26}$$

and there is

$$|b\rangle = \sum_{j=1}^{N} \beta_j |u_j^A\rangle . \tag{27}$$

Since

$$\langle b|b\rangle = \sum_{j=1}^{N} \beta_j^2 = 1 , \tag{28}$$

we have $\beta_j^2 \le 1,\ \ \forall j \in [1, N]$. In addition, since eigenvectors $|u_j^A\rangle$ for *j*=1,…, *N* are orthogonal, there is

$$\begin{cases} |u_j^A\rangle\langle u_j^A| \cdot |u_i^A\rangle\langle u_i^A| = \mathbf{0}_N , & \text{if } \ i \ne j \\ |u_j^A\rangle\langle u_j^A| \cdot |u_i^A\rangle\langle u_i^A| = |u_j^A\rangle\langle u_j^A| & \text{if } \ i = j \end{cases} . \tag{29}$$

Therefore, for matrix $AQ_bA$, there is

$$\begin{aligned} AQ_bA &= A(I_N - |b\rangle\langle b|)A \\ &= \left(\sum_{i=1}^{N} \lambda_i^A |u_i^A\rangle\langle u_i^A|\right) \cdot \left(I_N - \sum_{j=1}^{N}\sum_{k=1}^{N} \beta_j\beta_k |u_j^A\rangle\langle u_k^A|\right) \cdot \left(\sum_{l=1}^{N} \lambda_l^A |u_l^A\rangle\langle u_l^A|\right) \\ &= \left(\sum_{i=1}^{N} \lambda_i^A |u_i^A\rangle\langle u_i^A|\right) \cdot \left(\sum_{l=1}^{N} \lambda_l^A |u_l^A\rangle\langle u_l^A| - \sum_{j=1}^{N}\sum_{k=1}^{N}\sum_{l=1}^{N} \beta_j\beta_k\lambda_j^A |u_j^A\rangle\langle u_k^A|u_l^A\rangle\langle u_l^A|\right) \\ &= \left(\sum_{i=1}^{N} \lambda_i^A |u_i^A\rangle\langle u_i^A|\right) \cdot \left(\sum_{l=1}^{N} \lambda_l^A |u_l^A\rangle\langle u_l^A| - \sum_{l=1}^{N} \beta_l^2\lambda_l^A |u_l^A\rangle\langle u_l^A|\right) \\ &= \left(\sum_{i=1}^{N} \lambda_i^A |u_i^A\rangle\langle u_i^A|\right) \cdot \left(\sum_{l=1}^{N} \lambda_l^A (1-\beta_l^2) |u_l^A\rangle\langle u_l^A|\right) \\ &= \sum_{i=1}^{N} (\lambda_i^A)^2 (1-\beta_i^2) |u_i^A\rangle\langle u_i^A| = \sum_{i=1}^{N} \lambda_i^{AQ_bA} |u_i^A\rangle\langle u_i^A| \end{aligned} , \tag{30}$$

where $\lambda_i^{AQ_bA}$ represents the *i*-th eigenvalue of $AQ_bA$. Since $(\lambda_i^A)^2 \ge 0$ and $(1-\beta_i^2) \ge 0$, we have $\lambda_i^{AQ_bA} \ge 0$ (i.e., the eigenvalues of the matrix $AQ_bA$ are non-negative). For the F-norm of the Hermitian matrix $AQ_bA$, we have

$$\begin{aligned} \|AQ_bA\|_F^2 &= \sum_{i=1}^{N}\sum_{j=1}^{N} (AQ_bA)_{ij}^2 = Tr(AQ_bA \cdot (AQ_bA)^T) = \sum_{i=1}^{N} \lambda_i^{AQ_bA\cdot(AQ_bA)^T} \\ &\ge \lambda_{\max}^{AQ_bA\cdot(AQ_bA)^T} = \|AQ_bA\|_2^2 = (\lambda_{\max}^{AQ_bA})^2 \end{aligned} , \tag{31}$$

where $\lambda_{\max}^{X}$ represents the maximum eigenvalue of the matrix *X*. Since the eigenvalue of the matrix $AQ_bA$ is non-negative, there is $\|AQ_bA\|_F \ge \lambda_{\max}^{AQ_bA}$. Furthermore, scaling the matrix $AQ_bA$ by a coefficient will also proportionally scale its eigenvalues, such that:

$$0 \le \lambda_i^{AQ_bA/\|AQ_bA\|_F} = \frac{\lambda_i^{AQ_bA}}{\|AQ_bA\|_F} \le \frac{\lambda_i^{AQ_bA}}{\lambda_{\max}^{AQ_bA}} \le 1, \quad i = 1, \ldots, N \ . \tag{32}$$

According to Eq. (8), there is

$$\frac{AQ_bA}{\|AQ_bA\|_F} | x\rangle = \frac{AQ_b}{\|AQ_bA\|_F} | b\rangle = \boldsymbol{0}_N \ , \tag{33}$$

which means $| x\rangle$ is one of the eigenvectors of $AQ_bA / \|AQ_bA\|_F$, and the corresponding eigenvalue $\lambda_{\min}^{AQ_bA/\|AQ_bA\|_F}$ is 0.

For $AQ_bA / \|AQ_bA\|_F$, the eigenvalues $\lambda^{AQ_bA/\|AQ_bA\|_F}$ can be obtained by solving the following equation:

$$\frac{AQ_bA}{\|AQ_bA\|_F} - \lambda^{AQ_bA/\|AQ_bA\|_F} I_N = 0 \, . \tag{34}$$

For $H_1$, the eigenvalues $\lambda^{H_1}$ can be obtained by solving the following equation:

$$\begin{aligned} H_1 - \lambda^{H_1} I_N &= I_N - \frac{AQ_bA}{\|AQ_bA\|_F} - \lambda^{H_1} I_N \\ &= -\left( \frac{AQ_bA}{\|AQ_bA\|_F} - (1-\lambda^{H_1}) I_N \right) = 0 \end{aligned} \ . \tag{35}$$

According to Eqs. (34) and (35), we can find that the eigenvalues of $H_1$ satisfy

$$\lambda^{H_1} = 1 - \lambda^{AQ_bA/\|AQ_bA\|_F} \ , \tag{36}$$

and their corresponding eigenvectors are the same. In addition, since $\lambda_{\min}^{AQ_bA/\|AQ_bA\|_F} = 0$, there is

$$\lambda_{\max}^{H_1} = 1 - \lambda_{\min}^{AQ_bA/\|AQ_bA\|_F} = 1 \, , \tag{37}$$

and the corresponding eigenvector is $| x\rangle$.

In summary, the largest eigenvalue of $H_1$ is 1, with the corresponding eigenvector being $| x\rangle$, while the other eigenvalues belong to the interval [0, 1). The proof is complete.

Under this Hamiltonian design scheme, during the adiabatic evolution of the quantum system's Hamiltonian, the quantum state of the main qubits is maintained in the highest energy level (with the corresponding eigenvalue being 1) and slowly evolves from $| b\rangle$ to $| x\rangle$.

### *2.4. Eigenvalue Separator*

In a general adiabatic evolution process, when there are nearby energy levels close to the desired energy level, the quantum system may experience level transitions, leading to the failure of adiabatic evolution. For the Hamiltonians described in Sec. 2.3, this occurs when the final Hamiltonian has eigenvalues close to 1, ultimately compromising the quality of the solution or even causing the failure of the solution process. Therefore, we need to post-process the Hamiltonian to separate the target eigenvalue from other eigenvalues without altering the corresponding eigenvectors. The technique developed in this work for this post-processing of the Hamiltonian is referred to Eigenvalue Separator.

One of the key aspects of Eigenvalue Separator is the utilization of the matrix self-multiplication property. For the Hamiltonian $H_1$, there is

$$H_1 = \sum_{j=1}^{N} \lambda_j^{H_1} |u_j^A\rangle\langle u_j^A|. \tag{38}$$

For the matrix $H'_1 = H_1 \cdot H_1$, there is

$$\begin{aligned} H'_1 &= (\sum_{j=1}^{N} \lambda_j^{H_1} |u_j^A\rangle\langle u_j^A|) \cdot (\sum_{i=1}^{N} \lambda_i^{H_1} |u_i^A\rangle\langle u_i^A|) \\ &= \sum_{j=1}^{N} (\sum_{i=1}^{N} \lambda_j^{H_1} \lambda_i^{H_1} |u_j^A\rangle\langle u_j^A| \cdot |u_i^A\rangle\langle u_i^A|) \end{aligned}. \tag{39}$$

According to Eq. (29), there is

$$H'_1 = H_1 \cdot H_1 = \sum_{j=1}^{N} (\lambda_j^{H_1})^2 |u_j^A\rangle\langle u_j^A| \ . \tag{40}$$

It is evident that when a Hermitian matrix is multiplied by itself, the eigenvalues are squared while the corresponding eigenvectors remain unchanged.

The second key aspect of Eigenvalue Separator is ***Lemma 2*** mentioned in Sec. 2.3. For the final Hamiltonian $H_1$ in BEES-QDALS, since the eigenvalue corresponding to $|x\rangle$ is 1, squaring 1 still yields 1, which means that the eigenvalue of $H_1$ corresponding to the eigenvector $|x\rangle$ remains 1.When the matrix is multiplied by itself (resulting in $H'_1$), the other eigenvalues of $H_1$ belonging to the interval (0, 1) will tend to approach 0, while the eigenvalue 0 (if exists) remains unchanged. Extending this concept to multiple applications of matrix self-multiplication, as the number of self-multiplications *d* approaches positive infinity, there is

$$\lim_{d\to\infty} \mathrm{eig}(\prod_{i=1}^{d} H_1) = \{(\lambda_1^{H_1})^d, (\lambda_2^{H_1})^d, ..., (\lambda_N^{H_1})^d\} \approx \{0, ..., 0, (\lambda_{\max}^{H_1})^d = 1, 0, ..., 0\}, \tag{41}$$

where $\mathrm{eig}(\cdot)$ represents the set of eigenvalues of the Hamiltonian. It can be seen that as the number of self-multiplications of the matrix $H_1$ approaches positive infinity, the eigenvalues originally distributed in the interval [0, 1) tend toward zero, while the eigenvalue that was originally 1 remains unchanged. This means that through the self-multiplication of $H_1$, eigenvalue separation can be achieved for the final Hamiltonian in BEES-QDALS, while the corresponding eigenvectors remain unchanged.

Furthermore, we developed a *D*th-order Eigenvalue Separator, as illustrated in **Fig. 3**, which enhances the efficiency of eigenvalue separation through iterative self-multiplication. By utilizing the *D*th-order Eigenvalue Separator, $H_1$ is raised to the power of $2^D$, with the corresponding eigenvalues being

$$\mathrm{eig}((H_1)^{2^D}) = \{(\lambda_1^{H_1})^{2^D}, ..., (\lambda_{\max}^{H_1})^{2^D} = 1, ..., (\lambda_N^{H_1})^{2^D}\}. \tag{42}$$

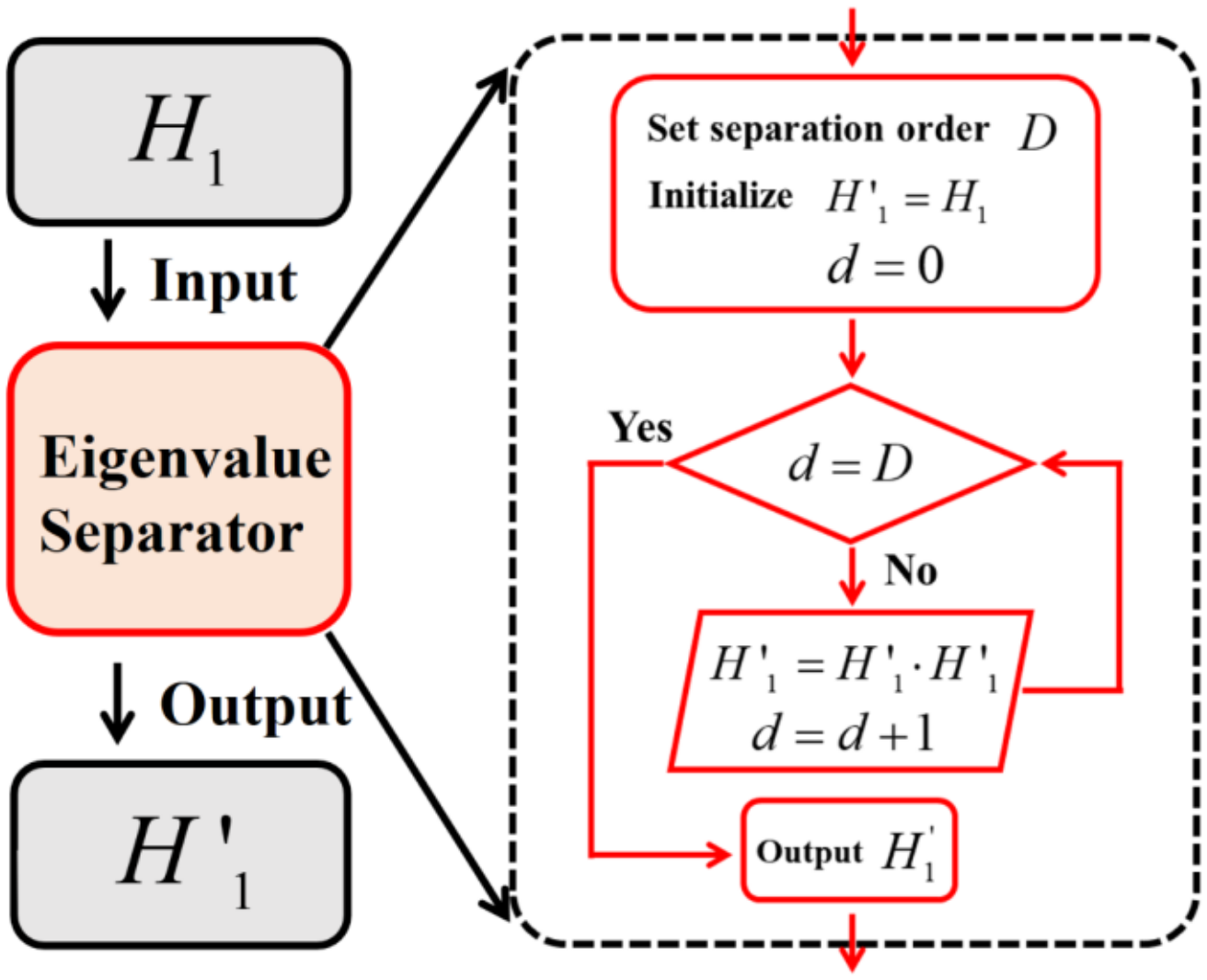

**Fig. 3.** Workflow of *D*th-order Eigenvalue Separator

### *2.5. Block encoding operator*

For the non-unitary matrix $H^{(1)}_{\exp,s}$ to be encoded, let $(H^{(1)}_{\exp,s})_{\propto}$ denote the matrix obtained by proportionally scaling all elements of $H^{(1)}_{\exp,s}$, there is $(H^{(1)}_{\exp,s})_{\propto} \propto H^{(1)}_{\exp,s}$. As shown in **Fig. 4**, through block encoding, the non-unitary matrix $(H^{(1)}_{\exp,s})_{\propto}$ is encoded into a block of a larger unitary matrix $U_H$ as $(\tilde{H}^{(1)}_{\exp,s})_{\propto}$ [26]. When the block encoding is sufficiently precise, we can obtain $\tilde{H}^{(1)}_{\exp,s} = H^{(1)}_{\exp,s}$, which allows for the equivalent implementation of a "non-unitary operation" on the quantum state through the post-selection of quantum states. In **Fig. 4**, $*$ denotes the arbitrary matrix elements. In this work, we refer to $H^{(1)}_{\exp,s}$ as the matrix to be encoded, and $\tilde{H}^{(1)}_{\exp,s}$ as the effective block within $U_H$.

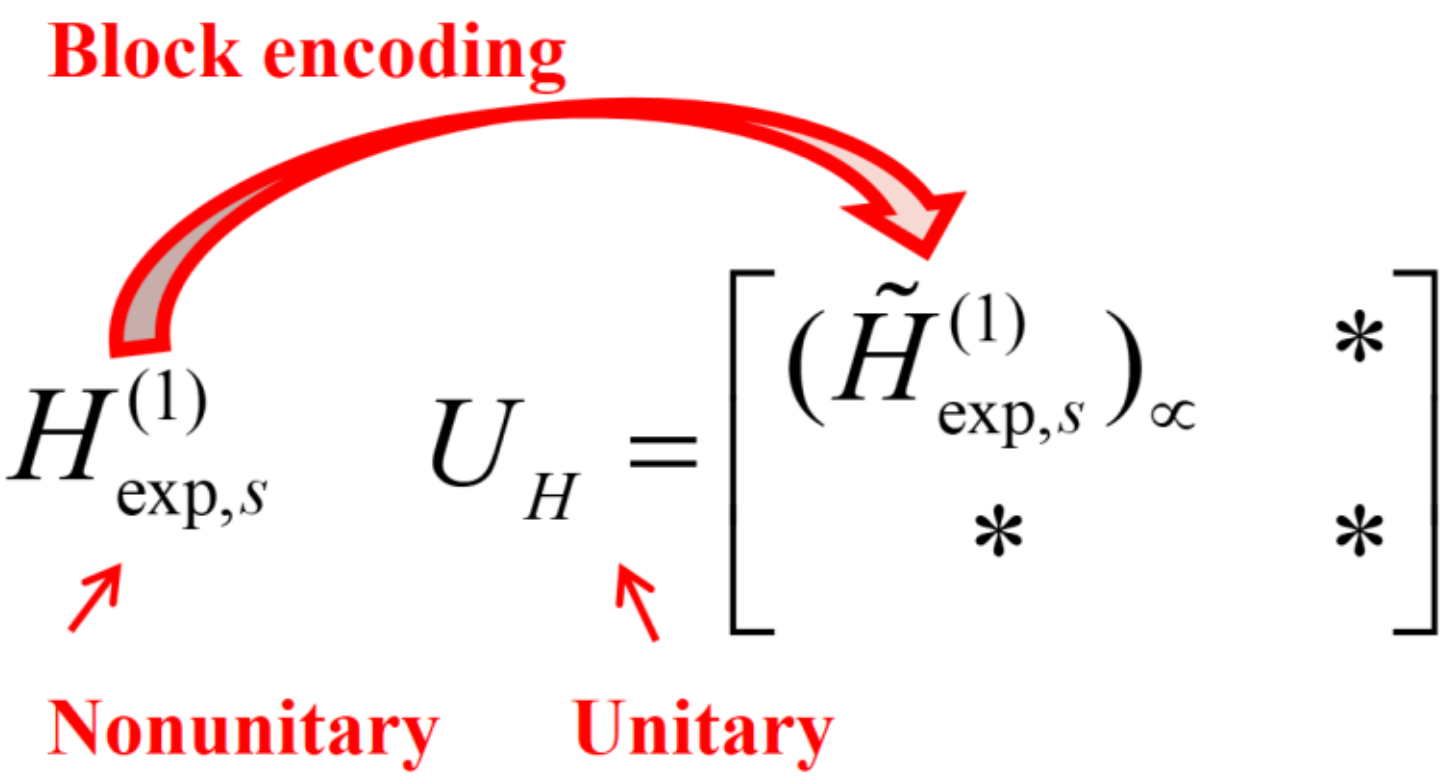

**Fig. 4.** Block encoding diagram

Let's start with a brief introduction to the original precise block encoding [27]. The universal precise block encoding quantum circuit is shown in **Fig. 5**, which is composed of three quantum registers $q_{TOP}$, $q_i$ and $q_j$, where $q_{TOP}$ and $q_i$ are $n_A = n+1 = \log_2 N + 1$ ancillary qubits, $q_j$ are $n = \log_2 N$ main qubits. The quantum register $q_{TOP}$ is used to store the matrix element values. In the original precise block encoding circuit, the quantum registers $q_i$ and $q_j$ are used to encode the row and column indices in binary form, respectively.

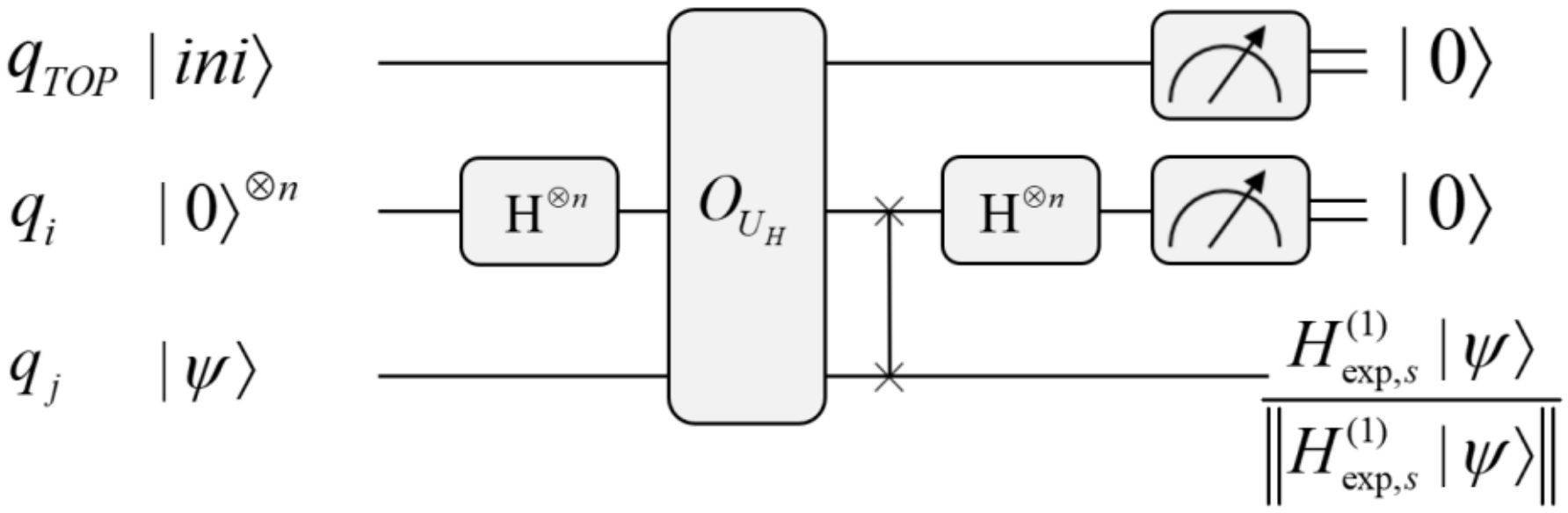

**Fig. 5.** Quantum circuit of universal precise block encoding

The core of matrix element encoding is the oracle $O_{U_H}$, as shown in **Fig. 6**, which consists of a series of controlled *RY* and controlled *RZ* gates. The control conditions for these control qubits are determined by the binary representation of the row and column indices of the currently encoded matrix element. The complex element of matrix $H^{(1)}_{\exp,s}$ at row *i*, column *j* is given by:

$$H^{(1)}_{\exp,s}[i,j] = h_{ij} = |h_{ij}| e^{\mathrm{i}\eta_{ij}} , \tag{43}$$

and the $O_{U_H}$ oracle is responsible for encoding the matrix element of $H^{(1)}_{\exp,s}$ into the amplitude of $q_{TOP}$ :

$$O_{U_H} |0\rangle |i\rangle |j\rangle = |h_{ij}\rangle |i\rangle |j\rangle , \tag{44}$$

with

$$|h_{ij}\rangle = |h_{ij}| e^{\mathrm{i}\eta_{ij}} |0\rangle + \sqrt{1-|h_{ij}|^2} e^{-\mathrm{i}\eta_{ij}} |1\rangle . \tag{45}$$

When $|ini\rangle = |0\rangle$, there is

$$RZ(2\phi_{ij}) RY(2\theta_{ij}) |0\rangle = \begin{bmatrix} e^{-i\phi_{ij}} & 0 \\ 0 & e^{i\phi_{ij}} \end{bmatrix} \begin{bmatrix} \cos(\theta_{ij}) & -\sin(\theta_{ij}) \\ \sin(\theta_{ij}) & \cos(\theta_{ij}) \end{bmatrix} \begin{bmatrix} 1 \\ 0 \end{bmatrix} = \begin{bmatrix} \cos(\theta_{ij}) e^{-i\phi_{ij}} \\ \sin(\theta_{ij}) e^{i\phi_{ij}} \end{bmatrix} . \tag{46}$$

According to Eq. (45), there is

$$RZ(2\phi_{ij}) RY(2\theta_{ij}) |0\rangle = \begin{bmatrix} |h_{ij}| e^{\mathrm{i}\eta_{ij}} \\ \sqrt{1-|h_{ij}|^2} e^{-\mathrm{i}\eta_{ij}} \end{bmatrix} . \tag{47}$$

According to Eqs. (46) and (47), the rotation parameters of *RZ* and *RY* gates are

$$\theta_{ij} = \arccos(|h_{ij}|) \tag{48}$$

$$\phi_{ij} = -\eta_{ij} . \tag{49}$$

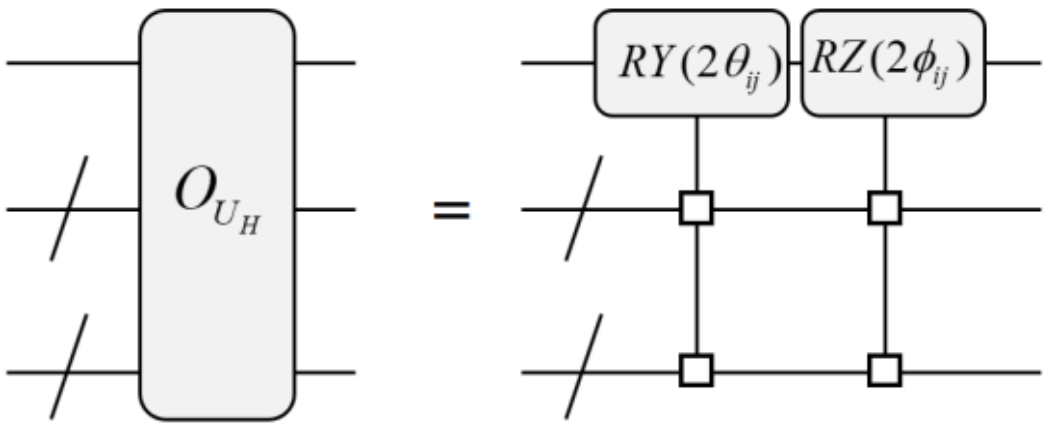

**Fig. 6.** Quantum circuit of $O_{U_H}$ oracle

The proper indexing of each elements in $H_{\exp,s}^{(1)}$ can be ensured by executing a series of *SWAP* gates.

It is worth noting that in the aforementioned original precise block encoding scheme, according to Eq. (48), even when encoding zero elements, the circuit still executes *RY* gates with $\theta_{ij} = \pi/2$. This means that quantum gates are required to encode zero elements. Since $H_{\exp,s}^{(1)}$ is typically sparse, with one of the most prominent characteristics of a sparse matrix being the presence of numerous zero elements, we develop an improved block encoding operator. In the new scheme, we initialize the state of the quantum register $q_{TOP}$ as $|ini\rangle = |1\rangle$ (from a circuit perspective, this is equivalent to executing an *X* gate on $q_{TOP}$ at the beginning). At this point, we have:

$$
\begin{aligned}
& RZ(2\tilde{\phi}_{ij})RY(2\tilde{\theta}_{ij})|1\rangle = \begin{bmatrix} e^{-i\tilde{\phi}_{ij}} & 0 \\ 0 & e^{i\tilde{\phi}_{ij}} \end{bmatrix} \begin{bmatrix} \cos(\tilde{\theta}_{ij}) & -\sin(\tilde{\theta}_{ij}) \\ \sin(\tilde{\theta}_{ij}) & \cos(\tilde{\theta}_{ij}) \end{bmatrix} \begin{bmatrix} 0 \\ 1 \end{bmatrix} \\
& = \begin{bmatrix} -\sin(\tilde{\theta}_{ij})e^{-i\tilde{\phi}_{ij}} \\ \cos(\tilde{\theta}_{ij})e^{i\tilde{\phi}_{ij}} \end{bmatrix} = \begin{bmatrix} |h_{ij}|e^{\mathrm{i}\eta_{ij}} \\ \sqrt{1-|h_{ij}|^2}e^{-\mathrm{i}\eta_{ij}} \end{bmatrix}
\end{aligned}, \tag{50}
$$

and the rotation parameters of *RY* and *RZ* gates are

$$\tilde{\theta}_{ij} = \arcsin(-|h_{ij}|), \tag{51}$$

$$\tilde{\phi}_{ij} = -\eta_{ij}, \tag{52}$$

which means when encoding zero elements, there is $\tilde{\theta}_{ij} = 0$ and $\tilde{\phi}_{ij} = 0$ (i.e., no longer require additional quantum gates), thus reducing the number of quantum gates in the block encoding circuit.

From the perspective of the entire block encoding process, for the system prior to evolution, there is:

$$|\psi_{\text{initial}}\rangle = \underbrace{|0\rangle|0\rangle^{\otimes \log_2 N}}_{n_A \text{ qubits}}|\psi\rangle = \begin{bmatrix} |\psi\rangle \\ \left.\begin{matrix} 0 \\ \vdots \\ 0 \end{matrix}\right\} 2^{n_A} \end{bmatrix} = |\psi, \boldsymbol{0}_{2^{n_A}}\rangle . \tag{53}$$

After applying the improved block encoding operator, the system state evolves to:

$$|\psi_{\text{final}}\rangle = U_H\left(X\underbrace{|0\rangle|0\rangle^{\otimes \log_2 N}}_{n_A \text{ qubits}}|\psi\rangle\right) = \begin{bmatrix} (H_{\exp,s}^{(1)}|\psi\rangle)_{\propto} \\ \left.\begin{matrix} * \\ \vdots \\ * \end{matrix}\right\} 2^{n_A} \end{bmatrix}, \tag{54}$$

with

$$(H^{(1)}_{\exp,s}\,|\psi\rangle)_{\propto} \propto H^{(1)}_{\exp,s}\,|\psi\rangle\,. \tag{55}$$

This means that through the post-selection operation, when the qubits in $q_{TOP}$ and $q_i$ collapse to $|0\rangle$, the state of the qubits in $q_j$ evolves to $H^{(1)}_{\exp,s}\,|\psi\rangle / \left\| H^{(1)}_{\exp,s}\,|\psi\rangle \right\|$, equivalently achieving a non-unitary operation on the quantum state.

It is noteworthy that the quantum state represented by the $2^{n_A}$ probability amplitudes marked with "*" in Eq. (54) is referred to as the "redundant state" (i.e., its information is useless). Due to the presence of redundant states, when continuous applications of the block encoding operator are required, it is necessary to "clear the redundant states" after each application, specifically:

$$(|\psi_{\text{final}}\rangle)_{CR} = \begin{bmatrix} \dfrac{H^{(1)}_{\exp,s}\,|\psi\rangle}{\left\| H^{(1)}_{\exp,s}\,|\psi\rangle \right\|} \\ \left.\begin{matrix} 0 \\ \vdots \\ 0 \end{matrix}\right\} 2^{n_A} \end{bmatrix}. \tag{56}$$

By the way, the specific block encoding circuit can be replaced with any more efficient approach without affecting the workflow of BEES-QDALS.

## 3. Case study

In the case study, we randomly generated a set $\mathcal{C} = \{C_2^1, C_2^2, C_4^1, C_4^2, C_8^1, C_8^2, C_{16}^1, C_{16}^2\}$ consisting of 8 complex linear equations and a set $\mathcal{S} = \{S_2^1, S_2^2, S_4^1, S_4^2, S_8^1, S_8^2, S_{16}^1, S_{16}^2\}$ composed of 8 sparse matrices. Here, $C$ represents the complex quantum linear system problem (QLSP), while $S$ denotes the sparse matrices, where the subscript indicates the dimension and the superscript denotes the sequence number. For instance, $C_2^1$ represents the first 2-dimensional complex QLSP, which is expressed as:

$$C_2^1: \quad \begin{bmatrix} 1.3088 + 0.0000i & 1.3246 - 0.6686i \\ 1.3246 + 0.6686i & 0.1441 + 0.0000i \end{bmatrix} |x\rangle \propto \begin{bmatrix} 0.7406 + 0.3019i \\ 0.4177 + 0.0914i \end{bmatrix}. \tag{57}$$

Taking $S_2^1$ as an example, it represents the first 2-dimensional sparse matrix

$$S_2^1 = \begin{bmatrix} 0 & 1.5912i \\ 0 & 0.5723 \end{bmatrix}. \tag{58}$$

In the following, Sec. 3.1 verifies the advantages of the improved block encoding over the original block encoding when applied to the sparse matrices in $\mathcal{S}$. Sec. 3.2 and Sec. 3.3 evaluate the impact of the newly designed Hamiltonian and the Eigenvalue Separator on the performance of HS-based QDALS and BE-based QDALS, respectively. Sec. 3.3 presents specific cases illustrating the eigenvalue separation process. Finally, Sec. 3.4 assesses the overall performance of BEES-QDALS.

Let $|\tilde{x}\rangle$ represent the result obtained from QDALS, and $|x\rangle$ denote the actual solution. We define the solution fidelity $F$ as

$$F = \langle \tilde{x} | x \rangle\,. \tag{59}$$

Since the primary indicators for characterizing the performance of QDALS are the solution fidelity $F$ and the required number of steps $M$, the case study mainly explores the effectiveness and superiority of BEES-QDALS and its components by comparing the fidelity $F$ when solving the QLSP in $\mathcal{C}$ at a fixed number of steps $M$, as well as the number of discrete adiabatic evolution steps $M$ needed to achieve the target fidelity $F_{tar}$ . Additionally, the robustness and superiority of BEES-QDALS are validated by statistically comparing the success rates of the original HS-based QDALS and BEES-QDALS in solving random QLSPs.

The quantum programming framework used in this study is PyQpanda, along with its default quantum simulator backend.

### *3.1. Validity verification of the improved block encoding and BE-based QDALS*

In this subsection, we validate the effectiveness and superiority of the improved block encoding operator. We perform block encoding on the matrices in $\mathcal{S}$ using both the original and improved block encoding methods. The matrix elements include pure real numbers, pure imaginary numbers, complex numbers, and zero elements, with zero elements accounting for approximately half of the total number of matrix elements (Due to the limited number of elements, the two-dimensional sparse matrix does not contain complex numbers). The errors and number of quantum gates required for both block encoding methods are compared in

**Table 1**. Here, the error is defined as the mean squared error (MSE) between elements of the effective block in the block encoding operator $U_H$ and the matrix to be encoded. "Saved" indicates the percentage of quantum gates saved by the improved block encoding operator compared to the original one.

The experimental results demonstrate that, when block encoding the matrices in $\mathcal{S}$ , both methods exhibit errors on the order of $10^{-15}$, essentially negligible. Meanwhile, the improved block encoding operator consistently uses fewer quantum gates. This implies that the improved method achieves precise block encoding while reducing the number of quantum gates. In the current case study, due to the presence of zero elements constituting around fifty percent of the matrix, the number of quantum gates is reduced by approximately fifty percent correspondingly. The more zero elements present, the greater the savings in quantum resources.

**Table 1.** The errors and number of quantum gates required for the original and improved block encoding schemes.

| Matrix | $S_2^1$ | | $S_2^2$ | | $S_4^1$ | | $S_4^2$ | |
|---|---|---|---|---|---|---|---|---|
| Number of zero elements | 2 | | 2 | | 9 | | 9 | |
| Method | Original | Improved | Original | Improved | Original | Improved | Original | Improved |
| MSE(1e-15) | 2.6174 | 1.3718 | 0.8421 | 1.1152 | 11.120 | 12.958 | 9.4222 | 6.2172 |
| Gates | 17 | 9 | 17 | 9 | 78 | 37 | 78 | 39 |
| Saved | 47.06% | | 47.06% | | 52.56% | | 50.00% | |

| Matrix | $S_8^1$ | | $S_8^2$ | | $S_{16}^1$ | | $S_{16}^2$ | |
|---|---|---|---|---|---|---|---|---|
| Number of zero elements | 33 | | 34 | | 127 | | 128 | |
| Method | Original | Improved | Original | Improved | Original | Improved | Original | Improved |
| MSE(1e-15) | 32.224 | 26.532 | 29.231 | 19.688 | 36.121 | 16.332 | 19.441 | 20.712 |
| Gates | 353 | 190 | 353 | 175 | 728 | 366 | 728 | 356 |
| Saved | 46.18% | | 50.42% | | 49.73% | | 51.10% | |

### *3.2. Validity verification of the new Hamiltonian*

In this subsection, we evaluate the performance of HS-based QDALS and BE-based QDALS under both the original and improved Hamiltonians to validate the accuracy and superiority of the new Hamiltonian. We utilize the Hamiltonian construction schemes from both the original QDALS framework and our improved approach to solve the Quantum Linear System Problem (QLSP) in $\mathcal{C}$ for a fixed number of steps *M*. The experimental results demonstrating the fidelity of the solutions for each Hamiltonian scheme are presented in **Table 2**. Specifically, O-HS and N-HS refer to the HS-based QDALS with the original and new Hamiltonians, respectively, while O-BE and N-BE correspond to the BE-based QDALS with the original and new Hamiltonians, respectively. The fidelities of solutions for the QLSP in $\mathcal{C}$ using HS-based QDALS and BE-based QDALS under both the original and new Hamiltonian schemes, with *M*=200, are depicted in **Fig. 7**.

**Table 2.** The fidelities of the solutions under the original and new Hamiltonian schemes when the number of steps *M* is fixed.

| Steps | Method | $C_2^1$ | $C_2^2$ | $C_4^1$ | $C_4^2$ | $C_8^1$ | $C_8^2$ | $C_{16}^1$ | $C_{16}^2$ |
|---|---|---|---|---|---|---|---|---|---|
| *M*=200 | O-HS | 0.9985 | 0.9596 | 0.9977 | 0.9485 | 0.7764 | 0.7958 | 0.3410 | 0.4148 |
| | N-HS | 0.9991 | 0.9994 | 0.9948 | 0.9923 | 0.9929 | 0.9931 | 0.6999 | 0.9875 |
| | O-BE | 0.0000 | 0.0000 | 0.0000 | 0.0000 | 0.0000 | 0.0000 | 0.0000 | 0.0000 |
| | N-BE | 1.0000 | 1.0000 | 0.9861 | 0.9921 | 0.7302 | 0.9645 | 0.1247 | 0.6077 |
| *M*=1000 | O-HS | 0.9992 | 1.0000 | 0.9988 | 0.9976 | 0.9169 | 0.9970 | 0.8894 | 0.9849 |
| | N-HS | 0.9990 | 0.9995 | 0.9995 | 0.9996 | 0.9985 | 0.9996 | 0.9799 | 0.9977 |
| | O-BE | 0.0000 | 0.0000 | 0.0000 | 0.0000 | 0.0000 | 0.0000 | 0.0000 | 0.0000 |
| | N-BE | 1.0000 | 1.0000 | 0.9994 | 0.9996 | 0.8058 | 0.9893 | 0.1509 | 0.7492 |
| *M*=2000 | O-HS | 1.0000 | 1.0000 | 0.9968 | 0.9998 | 0.9950 | 0.9660 | 0.8260 | 0.9965 |
| | N-HS | 0.9999 | 1.0000 | 0.9991 | 0.9998 | 0.9989 | 0.9984 | 0.9990 | 0.9992 |
| | O-BE | 0.0000 | 0.0000 | 0.0000 | 0.0000 | 0.0000 | 0.0000 | 0.0000 | 0.0000 |
| | N-BE | 1.0000 | 1.0000 | 0.9999 | 0.9999 | 0.8511 | 0.9945 | 0.1695 | 0.8175 |

For HS-based QDALS, when *M*=200, both O-HS and N-HS demonstrate high fidelity results for smaller linear systems (dimensions less than or equal to 4), with no significant difference between them. However, as the size of the linear system increases (i.e., dimensions greater than 4), the accuracy of N-HS shows a marked improvement over O-HS. This finding indicates that the Hamiltonian designed in this study enhances the fidelity of solutions obtained from HS-based QDALS.

In contrast, for BE-based QDALS with *M*=200, the solutions derived from the old Hamiltonian consistently fail. Upon integrating the new Hamiltonian, BE-based QDALS successfully solves the QLSP, highlighting its dependence on the newly designed Hamiltonian. Furthermore, for smaller linear systems, the fidelity of N-BE solutions is often comparable to that of N-HS. However, there remains significant potential for improvement in the fidelity of N-BE when addressing larger-scale QLSPs.

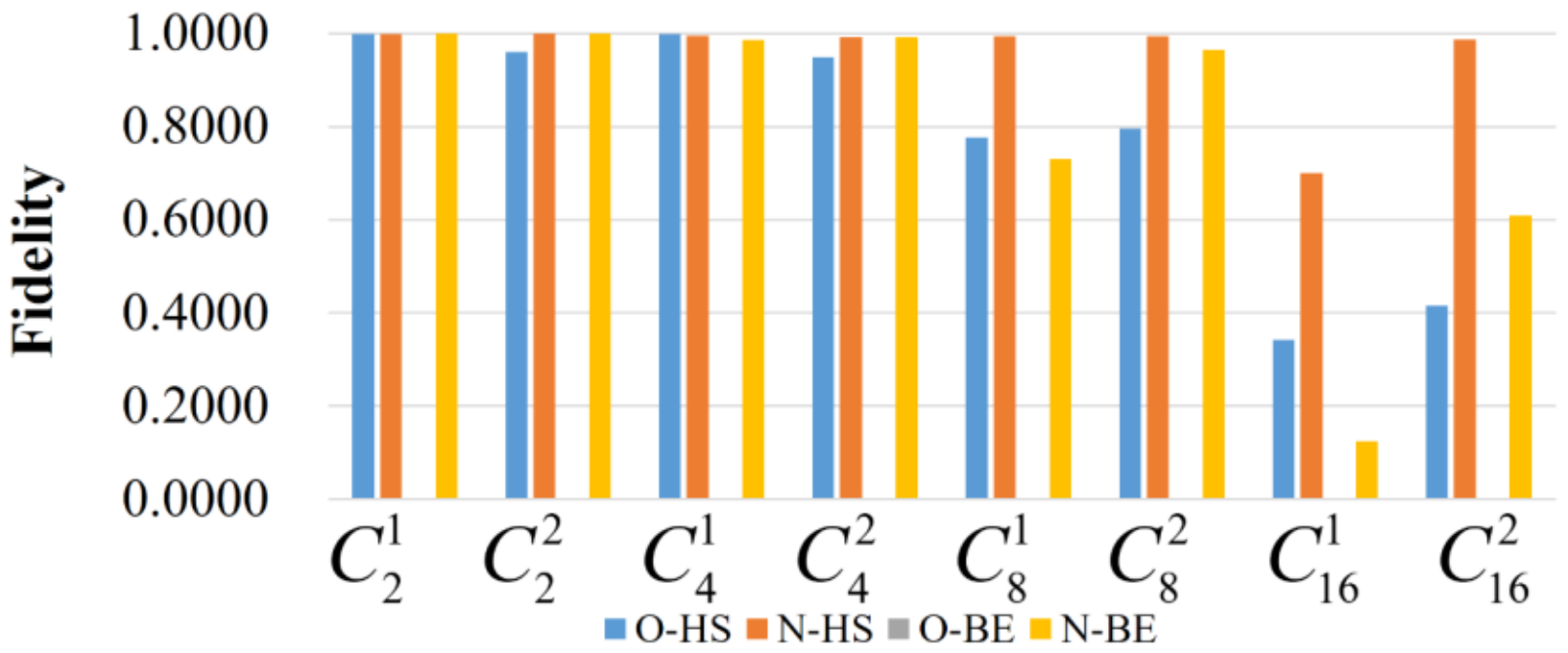

**Fig. 7.** The fidelities of the solution results obtained using the new and old Hamiltonians at $M$=200

Furthermore, the results shown in **Table 2** demonstrate a general trend for O-HS, N-HS, and N-BE: within a specific range, an increase in the number of discrete adiabatic evolution steps leads to higher fidelity in the QLSP solutions. Additionally, across the experimental range, the QDALS utilizing the new Hamiltonian consistently yields results with greater fidelity compared to those derived from the original Hamiltonian.

The experiments highlight the superiority of the QDALS with the new Hamiltonian in terms of solution accuracy at a fixed number of steps. Subsequently, we assessed the number of steps $M$ required to achieve the target fidelity of the solution. The experimental results are presented in **Table 3**, where "Failed" refers to instances where the target precision could not be attained within a finite number of steps. Specifically, when addressing the QLSP in $\mathcal{C}$ with a predefined target fidelity of $F_{tar}$=0.9999, the HS-based QDALS consistently accomplishes this with fewer evolution steps after employing the new Hamiltonian. In contrast, O-BE fails to reach the target precision within the limited number of steps, while N-BE completes the evolution within a finite number of steps.

**Table 3.** The number of steps required for each scheme to achieve a target fidelity of $F_{tar}$=0.9999.

| | Method | $C_2^1$ | $C_2^2$ | $C_4^1$ | $C_4^2$ | $C_8^1$ | $C_8^2$ | $C_{16}^1$ | $C_{16}^2$ |
|---|---|---|---|---|---|---|---|---|---|
| $F_{tar}$=0.9999 | O-HS | 33 | 24 | 274 | 264 | 1336 | 2482 | 5032 | 4668 |
| | N-HS | 10 | 18 | 255 | 248 | 1131 | 1841 | 3834 | 4487 |
| | O-BE | Failed | Failed | Failed | Failed | Failed | Failed | Failed | Failed |
| | N-BE | 15 | 28 | 2383 | 2124 | 258758 | 29876 | 303962 | 141860 |

In conclusion, the new Hamiltonian designed in this work significantly enhances the performance of QDALS. For HS-based QDALS, the use of the new Hamiltonian improves solution fidelity at a fixed number of steps, and it reduces the number of steps required to achieve the target fidelity when solving the QLSP. In the case of BE-based QDALS, the implementation of the new Hamiltonian facilitates successful QDALS solutions and enables the attainment of the target precision within a finite number of steps when a target fidelity is predefined.

### *3.3. Validity verification of the Eigenvalue Separator*

In this subsection, we begin by visualizing the 4th-order eigenvalue separation process of the final Hamiltonian $H_1$ when solving $C_4^1$. In this case, there is

$$H_1 = \begin{bmatrix} 0.8801 + 0.0000i & -0.0149 - 0.0557i & -0.0092 + 0.0280i & 0.0163 - 0.0544i \\ -0.0149 + 0.0557i & 0.3408 - 0.0000i & -0.2841 - 0.2685i & 0.0923 - 0.1583i \\ -0.0092 - 0.0280i & -0.2841 + 0.2685i & 0.6885 - 0.0000i & -0.0113 - 0.1592i \\ 0.0163 + 0.0544i & 0.0923 + 0.1583i & -0.0113 + 0.1592i & 0.8759 + 0.0000i \end{bmatrix}. \quad (60)$$

In the process of eigenvalue separation, the eigenvalues of $H'_1$ are shown in **Table 4** and **Fig. 8**. At a separation order of 0, the corresponding eigenvalues are the original four eigenvalues, with the maximum eigenvalue being 1 and the other eigenvalues distributed within the interval (0, 1). When the separation order is 1, the maximum eigenvalue, referred to as the target eigenvalue $\lambda_3^{H'_1}$, remains unchanged, while the neighboring eigenvalue $\lambda_4^{H'_1}$ decreases, resulting in an increased gap between $\lambda_3^{H'_1}$ and $\lambda_4^{H'_1}$. As the separation order continues to rise, this gap further widens, effectively achieving eigenvalue separation.

**Table 4.** The eigenvalues involved in the 4th-order eigenvalue separation process of $H'_1$.

| Separation order | 0 | 1 | 2 | 3 | 4 |
|---|---|---|---|---|---|
| $\lambda_1^{H'_1}$ | 0.0182 | 0.0003 | 0.0000 | 0.0000 | 0.0000 |
| $\lambda_2^{H'_1}$ | 0.8169 | 0.6673 | 0.4452 | 0.1982 | 0.0393 |
| $\lambda_3^{H'_1}$ | 1.0000 | 1.0000 | 1.0000 | 1.0000 | 1.0000 |
| $\lambda_4^{H'_1}$ | 0.9502 | 0.9029 | 0.8152 | 0.6645 | 0.4416 |

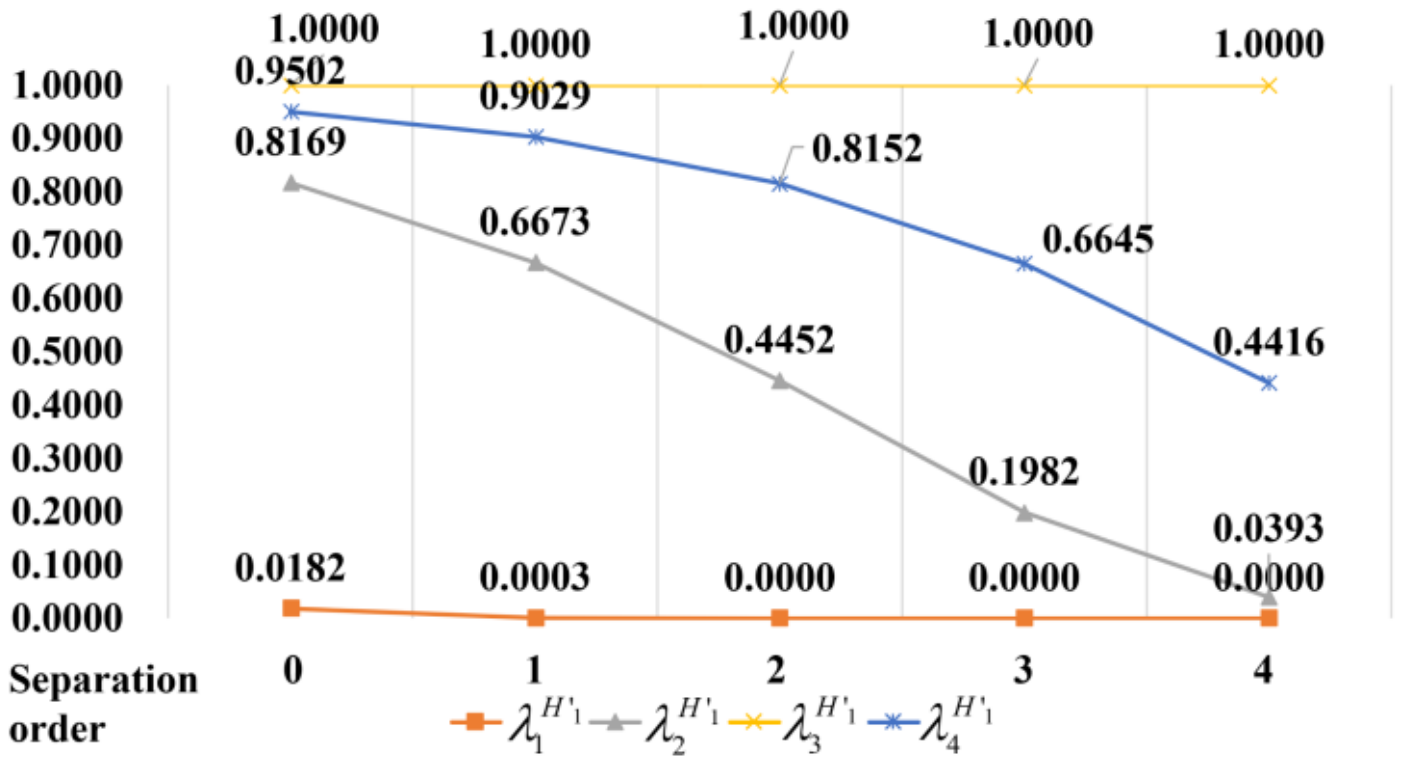

**Fig. 8.** Visualization of the 4th-order eigenvalue separation process for $H'_1$

The Eigenvalue Separator proposed in this work is based on the newly designed Hamiltonian herein. Therefore, we validate the superiority of the Eigenvalue Separator under the condition of utilizing the new Hamiltonian construction scheme. The fidelities of solving $C_{16}^1$ with different orders of eigenvalue separation are presented in **Table 5** and **Fig. 9**, where 0th-order eigenvalue separation is equivalent to not employing the eigenvalue separation module.

**Table 5.** The fidelities of the solutions for $C_{16}^{1}$ achieved through different orders of eigenvalue separation at a fixed number of steps *M*.

| Method | Separation Order | *M*=50 | *M*=100 | *M*=150 | *M*=200 |
|---|---|---|---|---|---|
| N-HS | 0 | 0.3185 | 0.5034 | 0.6325 | 0.6999 |
| | 2 | 0.9040 | 0.9929 | 0.9892 | 0.9777 |
| | 4 | 0.9000 | 0.9627 | 0.9368 | 0.9933 |
| | 6 | 0.6804 | 0.8795 | 0.9627 | 0.9831 |
| | 8 | 0.9872 | 0.9540 | 0.9224 | 0.9503 |
| | 10 | 0.9227 | 0.9277 | 0.9995 | 0.9764 |

| Method | Separation Order | *M*=50 | *M*=100 | *M*=150 | *M*=200 |
|---|---|---|---|---|---|
| N-BE | 0 | 0.1073 | 0.1160 | 0.1210 | 0.1247 |
| | 2 | 0.1372 | 0.1525 | 0.1645 | 0.1752 |
| | 4 | 0.2147 | 0.3071 | 0.4077 | 0.5134 |
| | 6 | 0.7807 | 0.9915 | 0.9996 | 1.0000 |
| | 8 | 1.0000 | 1.0000 | 1.0000 | 1.0000 |
| | 10 | 1.0000 | 1.0000 | 1.0000 | 1.0000 |

It is evident that at a fixed number of steps *M*, the fidelities of the results obtained by QDALS exhibit a positive correlation with the order of eigenvalue separation. The enhancement in fidelity is particularly notable in BE-based QDALS. Specifically, when the separation order is 6 or greater, N-BE consistently yields results with higher fidelity than N-HS. When the separation order reaches 8 or more, N-BE is capable of achieving nearly exact solutions at a fixed number of steps.

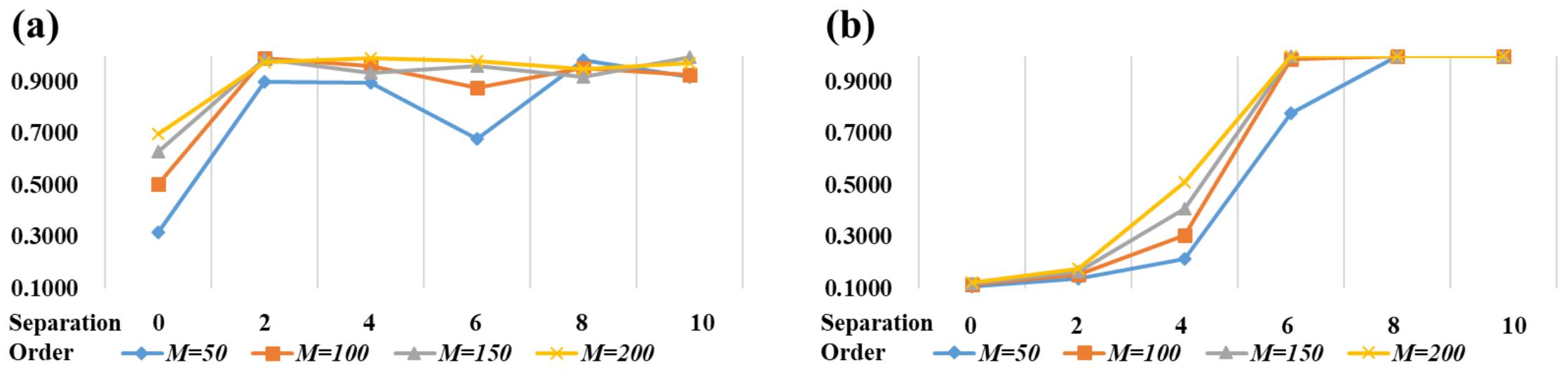

**Fig. 9.** Solution fidelities for $C_{16}^{1}$ with different separation orders at a fixed *M*:
(a) Results using the N-HS method, (b) Results using the N-BE method.

Next, we evaluate the number of steps *M* required to achieve the target fidelity of the solution, and the results are shown in **Table 6**, The findings reveal that increasing the separation order can significantly reduce the number of steps needed by QDALS. Additionally, compared with N-HS, N-BE exhibits a clear advantage at high target fidelities and high separation orders. Notably, once the separation order reaches 8, N-BE consistently outperforms N-HS across all metrics.

**Table 6.** The number of steps required to achieve the target fidelity  through different orders of eigenvalue separation.

| Method | Separation order | $F_{tar}$=0.9 | $F_{tar}$=0.99 | $F_{tar}$=0.999 | $F_{tar}$=0.9999 |
|---|---|---|---|---|---|
| N-HS | 0 | 497 | 1192 | 1954 | 3834 |
| | 2 | 37 | 100 | 327 | 2436 |
| | 4 | 31 | 80 | 313 | 2214 |
| | 6 | 25 | 67 | 246 | 1742 |
| | 8 | 11 | 56 | 178 | 1236 |
| | 10 | 9 | 39 | 150 | 504 |

| Method | Separation order | $F_{tar}$=0.9 | $F_{tar}$=0.99 | $F_{tar}$=0.999 | $F_{tar}$=0.9999 |
|---|---|---|---|---|---|
| N-BE | 0 | 75823 | 136465 | 204132 | 303962 |
| | 2 | 5202 | 9030 | 13795 | 19559 |
| | 4 | 451 | 750 | 1092 | 1482 |
| | 6 | 64 | 98 | 134 | 178 |
| | 8 | 7 | 21 | 27 | 34 |
| | 10 | 3 | 6 | 8 | 9 |

In conclusion, the Eigenvalue Separator proposed in this work significantly enhances the performance of QDALS. At a fixed number of steps, the implementation of the Eigenvalue Separator improves solution fidelity. To achieve the same target fidelity, fewer steps are required, with greater reductions in steps at higher eigenvalue separation orders. Notably, when the separation order exceeds 8, the performance of N-BE markedly surpasses that of N-HS.

### *3.4. Validity verification of the BEES-QDALS*

Following the validation of the performance enhancements of each component of BEES-QDALS on QDALS, we now proceed to a comprehensive performance comparison between the entire BEES-QDALS and the original QDALS. In this subsection, we employ BEES-QDALS with a separation order of 8 to solve the QLSP in $\mathcal{C}$ and compare the results with those obtained from the original QDALS, as illustrated in **Table 7**.

**Table 7.** The fidelities of the solutions obtained from the Original QDALS and BEES-QDALS

| Steps | Method | $C_2^1$ | $C_2^2$ | $C_4^1$ | $C_4^2$ | $C_8^1$ | $C_8^2$ | $C_{16}^1$ | $C_{16}^2$ |
|---|---|---|---|---|---|---|---|---|---|
| $M$=200 | Original QDALS | 0.9985 | 0.9596 | 0.9977 | 0.9485 | 0.7764 | 0.7958 | 0.3410 | 0.4148 |
| | BEES-QDALS | 1.0000 | 1.0000 | 1.0000 | 1.0000 | 1.0000 | 1.0000 | 1.0000 | 1.0000 |
| $M$=1000 | Original QDALS | 0.9992 | 1.0000 | 0.9988 | 0.9976 | 0.9169 | 0.9970 | 0.8894 | 0.9849 |
| | BEES-QDALS | 1.0000 | 1.0000 | 1.0000 | 1.0000 | 1.0000 | 1.0000 | 1.0000 | 1.0000 |
| $M$=2000 | Original QDALS | 1.0000 | 1.0000 | 0.9968 | 0.9998 | 0.9950 | 0.9660 | 0.8260 | 0.9965 |
| | BEES-QDALS | 1.0000 | 1.0000 | 1.0000 | 1.0000 | 1.0000 | 1.0000 | 1.0000 | 1.0000 |

The results demonstrate that, with four-decimal precision, the fidelities of the solutions obtained using BEES-QDALS approach 1. This indicates that BEES-QDALS is highly effective in solving the QLSP in $\mathcal{C}$, showing significant superiority in solution accuracy compared to the original QDALS. The number of steps $M$ required to reach the target fidelity is presented in **Table 8**. The results suggest that, for solving the QLSP in $\mathcal{C}$, when the target fidelity is set at $F_{tar}$=0.9999, BEES-QDALS also holds a clear advantage in terms of the number of steps required.

**Table 8.** The number of steps required to achieve the target fidelity of $F_{tar}$=0.9999 under the new and old Hamiltonian schemes.

| | Method | $C_2^1$ | $C_2^2$ | $C_4^1$ | $C_4^2$ | $C_8^1$ | $C_8^2$ | $C_{16}^1$ | $C_{16}^2$ |
|---|---|---|---|---|---|---|---|---|---|
| $F_{tar}$=0.9999 | Original QDALS | 33 | 24 | 274 | 264 | 1336 | 2482 | 5032 | 4668 |
| | BEES-QDALS | 1 | 1 | 1 | 1 | 18 | 4 | 34 | 13 |

Furthermore, we fix the number of steps at $M$=50,100,150,200 and use both the original QDALS and BEES-QDALS to solve 1000 randomly generated 16-dimensional complex QLSPs. We then categorize the results based on the number of solutions with fidelities in the ranges: less than 0.9, 0.9~0.99， 0.99~0.999， 0.999~0.9999, and 0.9999~1. The outcomes are presented in **Table 9**.

**Table 9.** At fixed step numbers of $M$=50,100,150,200, the statistical fidelity distribution of results obtained by Original QDALS and BEES-QDALS.

| Steps | Method | 0~0.9 | 0.9~0.99 | 0.99~0.999 | 0.999~0.9999 | 0.9999~1 |
|---|---|---|---|---|---|---|
| $M$=50 | Original QDALS | 979 | 21 | 0 | 0 | 0 |
| | BEES-QDALS | 33 | 59 | 54 | 66 | 788 |
| $M$=100 | Original QDALS | 972 | 28 | 0 | 0 | 0 |
| | BEES-QDALS | 20 | 17 | 32 | 41 | 890 |
| $M$=150 | Original QDALS | 948 | 51 | 1 | 0 | 0 |
| | BEES-QDALS | 13 | 11 | 29 | 24 | 923 |
| $M$=200 | Original QDALS | 895 | 101 | 4 | 0 | 0 |
| | BEES-QDALS | 5 | 18 | 12 | 15 | 950 |

A fidelity below 0.9 is defined as a solution failure, while a fidelity between 0.9999 and 1 is considered a complete success. The failure and complete success rates for both the original QDALS and BEES-QDALS in this case are presented in **Table 10**. As shown, across all fixed $M$, BEES-QDALS exhibits a significantly lower failure rate and a much higher complete success rate compared to the original QDALS, indicating a notable advantage.

**Table 10.** At fixed step numbers of $M$=50,100,150,200, the failure and complete success rates of original QDALS and BEES-QDALS.

| Steps | Method | Failure rate | Complete success rate |
|---|---|---|---|
| $M$=50 | Original QDALS | 97.9% | 0.0% |
| | BEES-QDALS | 3.3% | 78.8% |
| $M$=100 | Original QDALS | 97.2% | 0.0% |
| | BEES-QDALS | 2.0% | 89.0% |
| $M$=150 | Original QDALS | 94.8% | 0.0% |
| | BEES-QDALS | 1.3% | 92.3% |
| $M$=200 | Original QDALS | 89.5% | 0.0% |
| | BEES-QDALS | 0.5% | 95.0% |

In summary, BEES-QDALS demonstrates significantly stronger performance compared to the original QDALS. It achieves higher solution fidelity with the same number of steps and requires fewer steps to reach the same target fidelity.

## 4. Conclusion and future work

In this work, to reduce the complexity of block encoding operators, particularly concerning the number of required quantum gates, we have developed an improved block encoding operator that redesigns the block encoding circuit based on the properties of sparse matrices. This approach saves quantum gates used for encoding zero elements, thus reducing the overall usage of quantum gates while achieving precise block encoding. Furthermore, to circumvent the challenges associated with complex Hamiltonian simulations, we embed the improved block encoding technique into QDALS. By taking a first-order approximation of the Hamiltonian operator from the original QDALS, we implement equivalent non-unitary operations on qubits through the block encoding operator, enabling the solution of the Quantum Linear System Problem (QLSP) via discrete adiabatic evolution. Although the use of the first-order approximation and block encoding technique allows us to avoid complex Hamiltonian simulations, the errors introduced by the first-order approximation prevent the use of the original Hamiltonian. As a result, we design an improved Hamiltonian to ensure the feasibility of the BE-based QDALS. However, compared to the original QDALS (O-HS), the BE-based QDALS with the improved Hamiltonian (N-BE) requires an increased number of steps for discrete adiabatic evolution and struggles to achieve high-fidelity solutions. Therefore, we develop an Eigenvalue Separator to increase the gap between the eigenvalues of the Hamiltonian, ultimately enabling solutions to the QLSP with fewer steps and higher precision. Based on the improved block encoding operator and Eigenvalue Separator, we design a new QDALS, named BEES-QDALS. BEES-QDALS demonstrates significantly better performance compared to the original QDALS.

It is worth noting that the current design of the improved block encoding focuses primarily on optimizing the number of quantum gates. Exploring optimizations regarding the number of qubits to reduce the complexity of quantum circuits may become a future research direction. Additionally, the order of the Eigenvalue Separator is a predefined hyperparameter, and in the future, we could adaptively terminate the eigenvalue separation process based on the rate of change of matrix elements, thereby improving computational efficiency.

## Acknowledgments

This work was supported by the National Key Research and Development Program of China under Grant. 2022YFB3304700, the Open Research Project of the State Key Laboratory of Industrial Control Technology under Grant. ICT2024B21, and the National Natural Science Foundation of China under Grant. 62273016.